\documentclass{aastex631}

\usepackage{amsmath,nccmath}
\usepackage{mathrsfs}

\newcommand{\fract}[2]{\leavevmode\kern.1em
          \raise.5ex\hbox{\the\scriptfont0 #1}\kern-.1em
    \raise.15ex\hbox{\the\scriptfont0 /}\kern-.08em\lower.25ex\hbox{\the\scriptfont0 #2}}

\newcommand{\kps}{$\rm km\,s^{-1}$}

\newcommand{\half}{{\textstyle\frac{1}{2}}}

\newcommand{\quart}{{\textstyle\frac{1}{4}}}
\newcommand{\re}{\mathop{\rm Re}\nolimits}
\newcommand{\im}{\mathop{\rm Im}\nolimits}

\newcommand{\pderiv}[2]{\frac{\partial#1}{\partial#2}}

\renewcommand{\a}{{\boldsymbol{a}}}
\renewcommand{\b}{{\boldsymbol{b}}}

\renewcommand{\k}{\boldsymbol{k}}

\newcommand{\boldv}{{\boldsymbol{v}}}

\newcommand{\B}{{\boldsymbol{B}}}
\newcommand{\C}{{\boldsymbol{C}}}

\newcommand{\F}{{\boldsymbol{F}}}
\newcommand{\I}{{\mathrm{I}}}
\newcommand{\J}{{\mathrm{J}}}
\renewcommand{\S}{{\mathrm{S}}}
\newcommand{\Q}{{\mathrm{Q}}}

\newcommand{\X}{\boldsymbol{X}}
\newcommand{\Y}{\boldsymbol{Y}}
\newcommand{\M}{{\mathrm{M}}}

\newcommand{\U}{{\mathrm{U}}}

\renewcommand{\P}{{\mathrm{P}}}

\newcommand{\V}{\mathbf{V}}

\newcommand{\rhoc}{\rho_{\mathrm{c}}}
\newcommand{\rhon}{\rho_{\mathrm{n}}}
\newcommand{\cc}{c_{\mathrm{c}}}
\newcommand{\ac}{a_{\mathrm{c}}}
\newcommand{\cn}{c_{\mathrm{n}}}

\newcommand{\boldvc}{{\boldsymbol{v}_{\mathrm{c}}}}
\newcommand{\boldvn}{{\boldsymbol{v}_{\mathrm{n}}}}

\newcommand{\nunc}{\nu_{\mathrm{nc}}}

\newcommand{\nucn}{\nu_{\mathrm{cn}}}
\newcommand{\acn}{\alpha_{\mathrm{cn}}}

\newcommand{\R}{\mathrm{R}}

\newcommand{\vdot}{{\boldsymbol{\cdot}}}
\newcommand{\vcross}{{\boldsymbol{\times}}}
\newcommand{\grad}{\mbox{\boldmath$\nabla$}}

\newcommand{\diag}{\mathop{\rm diag}}
\newcommand{\thth}{\hspace{1.5pt}}
\newcommand{\curl}{\grad\vcross}

\newcommand\Div{\grad\vdot\thth}

  \renewcommand{\le}{\leqslant}

\begin{document}

\title{Efficiency of MHD Wave Generation in Weakly Ionized Atmospheres}

\author[0000-0001-5794-8810]{Paul S. Cally}
\affiliation{School of Mathematics, Monash University \\
Victoria 3800, Australia}

\begin{abstract}
Generation of Alfvén and slow magneto-acoustic waves in weakly ionized atmospheres by excitation of the charges-only component of the two fluid (charges and neutrals) plasma is shown to be more or less efficient depending on the energy fraction initially allocated to the three stationary ``flow differential'' modes which characterize the inter-species drift. This is explained via detailed analysis of the full ten-dimensional spectral description of two-fluid linear magneto\-hydro\-dynamics. Excitation via the velocity of the charges only is found to be very inefficient, in accord with previous results, whilst excitation via the magnetic field perturbation alone is highly efficient. All ten eigenvalues and eigenvectors are presented analytically in the high collision frequency regime.
\end{abstract}



\section{Introduction} \label{sec:intro}

The low atmospheres of cool stars are generally very weakly ionized. For example, the quiet solar photosphere has an ionization fraction as low as $10^{-4}$ \citep{KhoColDia14aa}. This has led to some discussion about whether Alfv\'en waves can be excited there. Alfv\'en waves are often invoked as important drivers of the solar wind and coronal heating \citep{Cravan05aa,McIde-Car11aa}, so this is an important issue.

To date, the argument has revolved around timescales, and less obviously the wave initialization mechanism. The electron-ion elastic scattering collision frequency at the quiet Sun photospheric base is of order $10^{10}$ $\rm s^{-1}$, which allows the charges to be modelled as a single fluid at frequencies much lower than this. The coupling between the neutrals fluid and the charges fluid is then described by the charges-neutrals collision frequency $\nucn$ of around $10^9$ $\rm s^{-1}$ and the neutral-charges collision frequency $\nunc$ of about $10^6$ $\rm s^{-1}$, dropping to a few tens of thousands per second at the top of the photosphere. The two fluids (neutrals and charges) are coupled only via these collisions.

The standard one-fluid (1F) ideal magneto\-hydro\-dynamic (MHD) formula for Alfv\'en wave energy flux density (energy per unit area per unit time) is
\begin{equation}\label{F}
F=E a
\end{equation}
directed along the field lines, where $E=E_\mathrm{kin}+E_\mathrm{mag}$ is the wave energy density made up equally of kinetic and magnetic contributions, and $a$ is the Alfv\'en speed. In terms of the plasma velocity amplitude $v$ and equilibrium density $\rho$, $E_\mathrm{mag}=E_\mathrm{kin}=\quart\rho v^2$. The extra factor of $\half$ comes from RMS averaging. Thus $F=\half\rho v^2\,a$ overall.
\citet{VraPoePan08aa} argued that this should be reduced by the factor $\chi^{-2}$ in a two-fluid (2F, charges and neutrals) plasma, where the ratio of neutrals to charges density $\chi=\rhon/\rhoc$ is typically of order $10^3-10^4$. This is because any velocity originally on the charges alone must quickly be shared with the neutrals, thereby greatly reducing $v$. Based on this insight, they concluded that a significant flux of Alfv\'en waves could not be generated in the solar photosphere.

On the other hand, \citet{TsaSteKop11aa} found no such diminution of flux in their 2F model. The cause of the discrepancy was discussed by \citet{SolCarBal13aa}, who concluded that the issue hinges on whether or not the neutrals fluid is given initial velocity $\boldvn$ matching that of the charges fluid $\boldvc$. If it is, then there is no collisional quenching of the velocity and the full 1F energy flux formula applies. If on the other hand the neutrals are left unperturbed ($\boldvn=\mathbf{0}$) by the driver, then the $\chi^{-2}$ reduction must be applied. \citeauthor{SolCarBal13aa}~judged that both viewpoints are valid; it just depends upon the nature of the perturbation that initiates the wave. 

The nature of the dependence on initial conditions is explored further here using a spectral description.

\subsection{Spectral Overview} \label{sec:overview}
For specified wavevector $\k$, it is shown that the linearized system is fully described by a tenth order eigenvalue system, corresponding to six slowly decaying MHD wave modes (slow, Alfv\'en and fast, each propagating in the $\pm\k$ directions); one stationary very slowly decaying isobaric mode with zero total pressure; and three very rapidly decaying (nanoseconds) stationary flow differential modes whose main characteristic is that they exhibit very different velocities of the neutrals and charges fluids. Only the MHD modes transport energy, and only the flow differential modes have significantly discrepant velocities between the two species. The eigenmodes are independent of each other, so their energies are simply whatever they were initially given by the excitation process, subject to their respective fast or slow collisional decays.

Any energy given to the flow differential modes is immediately lost to collisions, and represents an inefficiency of the excitation. For many simple excitations, this is around 50\%. Energy deposited in the isobaric mode also does not contribute to energy flux, but typically very little of this mode is generated. Energy density $E_i$ given to each MHD mode (where $i$ represents one of the six modes) plays a full role in energy transport, carrying flux $\F_i=E_i \V_i$, where $\V_i$ is the corresponding group velocity for that mode. Typically, realistic excitation results in nearly equal energies in the positively and negatively directed versions of each mode type, and so \emph{net} flux is small or zero, but flux in each direction can independently be large.

However, pure excitation via just a magnetic field perturbation does not place significant energy in the flow differential modes, and so essentially all energy is efficiently allocated to travelling MHD modes, 50\% in each direction. Therefore, a purely magnetic excitation can in fact be a most efficient generator of MHD waves. The same holds true for full-plasma excitations where charges and neutrals fluids are given the same initial velocities, as in \citet{TsaSteKop11aa}, because again the flow differential modes are not significantly excited.

The eigensystem decomposition is described in detail in Section \ref{sec:math}. Section \ref{sec:Energy} discusses the quadratic expressions for wave energy and flux. An analytic analysis of three different charges-only Alfv\'en wave excitation mechanisms is presented in Section \ref{sec:Alf}. Numerical examples of both Alfv\'en and magneto-acoustic excitation corresponding to different levels in the low solar atmosphere are set out in Section \ref{sec:res}. The results are discussed and summarized in Section \ref{sec:disc}.

\section{Mathematical Formulation} \label{sec:math}
\subsection{Governing Equations}\label{sec:eqns}
Consider a partially ionized hydrogen plasma. Extension to a more realistic chemical mixture does not change the arguments to be presented. The linearized coupled two-fluid (2F) wave equations take the form \citep{SolCarBal13gq}
\begin{subequations}\label{basiceqns}
\begin{gather}
\rhon \pderiv{\boldvn}{t}=-\grad p_{\mathrm{n}}+\acn(\boldvc-\boldvn),\label{mmntm n}\\[4pt]
  \pderiv{p_{\mathrm{n}}}{t}=-\gamma P_{\mathrm{n}} \Div\boldvn, \label{p n}\\[4pt]
  \intertext{for the neutrals, and}
  \rhoc \pderiv{\boldvc}{t}=-\grad p_{\mathrm{c}}+\frac{1}{\mu}(\curl\b)\vcross\B-\acn(\boldvc-\boldvn),\label{mmntm c}\\[4pt]
    \pderiv{p_{\mathrm{c}}}{t}=-\gamma P_{\mathrm{c}} \Div\boldvc,\label{p c}\\[4pt]
     \pderiv{\b}{t}=\curl(\boldvc\vcross\B),\label{b}
\end{gather}
\end{subequations}
for the charges, where the equilibrium magnetic field $\B$ and charges and neutrals gas pressures $P_{\mathrm{c}}$ and $P_{\mathrm{n}}$ are perturbed respectively by $\b$, $p_{\mathrm{c}}$ and $p_{\mathrm{n}}$. These equations follow from linearizing Equations (1) of \citet{PopLukKho19aa} under the assumption of temperature equality between charges and neutrals. 
The equilibrium charges and neutrals mass densities are $\rhoc$ and $\rhon$, $\acn$ is the friction coefficient, and $\mu$ is the vacuum permeability. The fluid velocities of the charges and neutrals are respectively $\boldvc=(u_{\mathrm{c}},v_{\mathrm{c}},w_{\mathrm{c}})$ and $\boldvn=(u_{\mathrm{n}},v_{\mathrm{n}},w_{\mathrm{n}})$. Without loss of generality, $\B$ is arbitrarily oriented in the $z$-direction and the wavevector $\k\equiv-i\,\grad=(\sin\theta,0,\cos\theta)k$ in the $x$-$z$ plane. Since $\b$ is perpendicular to $\k$ (which also follows from $\Div\b=0$), we need only retain two components, $b_y$ and $b_\perp=b_x\cos\theta-b_z\sin\theta$.

These 2F energy equations above are adiabatic, and do not explicitly feed energy lost via collisions back into the thermal state of the plasma. However, it is to be understood that this is where it goes \citep{CalGom23aa}.

Although Equations (\ref{basiceqns}) are commonly used in two-fluid studies, \citet{VraPoePan08aa} \citep[and more recently][]{AlhBalFed22aa} questioned the significance of the Lorentz force term on the right hand side of Equation (\ref{mmntm c}) on the basis that the collisional frequency greatly exceeds the ion gyrofrequency $\Omega_\mathrm{i}$ and hence that ions traverse only a tiny portion of their gyration path before scattering. However, \citet{TsaSteKop11aa} demonstrate that the ratio of the Lorentz force to the net ion-neutral collision force is of order 
\begin{equation*}
 \frac{\Omega_\mathrm{i}}{\nu_\mathrm{in}} : \frac{|\boldvc-\boldvn|}{|\boldvc|}.
\end{equation*}
The left hand side is typically $10^{-3}-10^{-2}$ in the low solar atmosphere \citep[see][Fig.~1]{KhoColDia14aa}. However, rapid collisions easily ensure that the right hand side is smaller than this, so the Lorentz force dominates or is at least comparable to the collisional force. Indeed, as will be shown, the collisions impose two timescales on the system: (i) a very rapid scale of order nanoseconds over which $|\boldvc-\boldvn|$ is reduced almost to zero, and (ii) a much longer ambipolar diffusion timescale which typically exceeds Alfv\'en wave periods of interest. The Lorentz force is unimportant over the former scale, but clearly significant over the latter, so it is retained here.

\subsection{Modal Eigenfrequencies and Eigenvectors}\label{sec:modal}

Let $\X=(u_{\mathrm{n}},v_{\mathrm{n}},w_{\mathrm{n}}, \psi_{\mathrm{n}}=p_{\mathrm{n}}/\rhon,u_{\mathrm{c}},v_{\mathrm{c}},w_{\mathrm{c}}, \psi_{\mathrm{c}}=p_{\mathrm{c}}/\rhoc, \alpha_\perp=b_\perp/\sqrt{\mu\,\rho}, \alpha_y=b_y/\sqrt{\mu\,\rho})^T$, where $\rho=\rhoc+\rhon$ is the total equilibrium density. Then Equations (\ref{basiceqns}) can be written in matrix form $d\X/dt=\M\,\X$ with exact solution $\X(t)=e^{\M\,t}\X(0)$. The matrix $\M$ is not defective so the matrix exponential may be constructed by direct exponentiation of the eigenvalues,  $e^{\M\,t}=\P^T \diag[e^{\lambda_{1}t},\ldots,e^{\lambda_{10}t}]\P^{-1}$, where the $\lambda_m$ are the eigenvalues of $\M$ and the columns of the $10\times10$ matrix $\P$ are the corresponding eigenvectors. 

The following quantities are introduced to define $\M$: neutral to charges mass ratio $\chi=\rhon/\rhoc$, (total) Alfv\'en velocity $\a=\B/\sqrt{\mu\rho}$, and (total) sound speed $c$ defined in terms of the charges and neutrals squared sound speeds $\cc^2=\gamma P_{\mathrm{c}}/\rhoc$ and $\cn^2=\gamma P_{\mathrm{n}}/\rhon$ by $c^2=(\cc^2+\chi\cn^2)/(1+\chi)$. Thermal equilibrium is assumed between the species, $\cc^2=2\cn^2$, so $c^2=\cn^2(2+\chi)/(1+\chi)$.  The total Alfv\'en speed $a=B/\sqrt{\mu\rho}$ is related to the charges-only Alfv\'en speed $\ac=B/\sqrt{\mu\rhoc}$ by $\ac=a\sqrt{1+\chi}$. It is natural to use $a$ rather than $a_\text{c}$ as the measure of magnetic influence in the strongly coupled regime. We also introduce the charges-neutral collision frequency $\nucn=\acn/\rhoc$ and the neutral-charges collision frequency $\nunc=\acn/\rhon$, related by $\nucn=\chi\nunc$.

The matrix $\M$ is written down explicitly in Appendix \ref{app:M}. In the absence of collisions, the top left $4\times4$ and bottom right $6\times6$ submatrices of $\M$ decouple, yielding two oppositely directed acoustic waves and two steady (zero frequency) incompressive velocity shears on the neutrals, and the six oppositely directed slow, Alfv\'en and fast waves on the charges.

On the other hand, the ten eigenfrequencies $\omega=i\,\lambda$ and the accompanying eigenvectors of the full collisionally coupled system are quite different in nature. All eigenfrequencies are complex for finite non-zero $\nunc$ and may be calculated numerically for any combination of parameters. It is instructive to calculate their asymptotic values in the large collision rate regime $\nunc\gg\omega$ that applies to waves of interest in the low solar atmosphere. 

A perturbation method for deriving these formulae is sketched in Appendix \ref{app:pert}.
\begin{subequations}\label{omegaAs}

\paragraph{Alfv\'en waves}
\begin{equation}
\omega_\text{A}=a\, k \cos\theta\left(\pm 1 - \frac{i\,a\,k\,\chi\cos\theta}{2(1+\chi)\nunc} 
+ \mathcal{O}\left((\omega/\nunc)^2\right) \right),  \label{AlfAs}
\end{equation}
which accords with Equation (21) of \citet{de-Hae98aa} for the two-dimensional case $\theta=0$. The corresponding eigenvectors are
\begin{equation}
A_\pm=\left(0,\mp 1-\frac{i\,a\,k_z(\chi+2)}{2(\chi+1)\nunc}+\mathcal{O}\left(\nunc^{-2}\right),0,0,0,\mp 1+\frac{i\,a\,k_z\,\chi}{2(\chi+1)\nunc}+\mathcal{O}\left(\nunc^{-2}\right),0,0,0,1\right), \label{AlfSlip}
\end{equation}
which verifies that the velocities on the neutrals and charges (second and sixth components) differ only at $\mathcal{O}(\nunc^{-1})$.

\paragraph{Slow waves}
\begin{equation}
\omega_\text{s}=\pm\frac{k \sqrt{a^2+c^2-\Delta }}{\sqrt{2}}
-\frac{i k^2 \chi  H_\text{s}}{2 \nunc (\chi +1) (\chi +2)^2 \left(a^2+c^2-\Delta\right)\Delta }
   +O\left((\omega/\nunc)^2\right), \label{slowAs}
\end{equation}
where $\Delta =\sqrt{a^4+c^4-2a^2c^2\cos2\theta}
$ satisfies $ |c^2-a^2|\le\Delta\le a^2+c^2$, and 
\begin{multline}
H_\text{s}=a^2 c^2 \cos 2 \theta  \left(a^2 (\chi +2) (\chi +3)+c^2 (\chi +3)-\Delta  (\chi +2)\right)+\Delta  \left(a^4 (\chi +2)^2+a^2
   c^2 (\chi +2)+c^4\right)\\
   -\left(a^2 (\chi +2)+c^2\right) \left(a^4 (\chi +2)+c^4\right).
\end{multline}
The corresponding eigenvector up to and including terms of $\mathcal{O}(\nunc^{-1})$ is far too long to present here, so we give it to leading order only. For the positively (upper sign) and negatively (lower sign)  directed modes, the eigenvector is
\begin{multline}\label{Xslow}
S_\pm=\left(\mp\frac{\sqrt{a^2+c^2-\Delta }}{\sqrt{2} a},0,\pm\frac{\cot \theta  \left(a^2-c^2+\Delta \right)}{\sqrt{2} a \sqrt{a^2+c^2-\Delta }},\frac{c^2 (\chi
   +1) \csc \theta  \left(a^2 \cos 2 \theta -c^2+\Delta \right)}{a (\chi +2) \left(a^2+c^2-\Delta \right)},\mp\frac{\sqrt{a^2+c^2-\Delta }}{\sqrt{2}
   a},0,\right.\\
   \left. \pm\frac{\cot \theta  \left(a^2-c^2+\Delta \right)}{\sqrt{2} a \sqrt{a^2+c^2-\Delta }},\frac{2 c^2 (\chi +1) \csc \theta  \left(a^2 \cos 2 \theta
   -c^2+\Delta \right)}{a (\chi +2) \left(a^2+c^2-\Delta \right)},1,0\right) +\mathcal{O}\left(\nunc^{-1}\right).
\end{multline}
Again the velocities on the neutrals and charges differ only at $\mathcal{O}(\nunc^{-1})$.

\paragraph{Fast waves}
\begin{equation}
\omega_\text{f}=\pm\frac{k \sqrt{a^2+c^2+\Delta }}{\sqrt{2}}-\frac{i k^2 \chi H_\text{f}}{2 \nunc (\chi +1) (\chi +2)^2 \left(a^2+c^2+\Delta\right)\Delta }  +O\left((\omega/\nunc)^2\right), \label{fastAs}
\end{equation}
where
\begin{multline}
H_\text{f}=-a^2 c^2 \cos 2 \theta  \left(a^2 (\chi +2) (\chi +3)+c^2 (\chi +3)+\Delta  (\chi +2)\right)+\Delta  \left(a^4 (\chi +2)^2+a^2
   c^2 (\chi +2)+c^4\right)\\
   +\left(a^2 (\chi +2)+c^2\right) \left(a^4 (\chi +2)+c^4\right).
\end{multline}
The eigenvector $F_\pm$ for the positively directed fast mode is exactly as for $S_\pm$ given in Equation (\ref{Xslow}) but with $\Delta$ replaced by $-\Delta$.

\paragraph{Flow differential mode rapid decay (triple)}
\begin{equation}
\omega_\text{diff}=-i\,(1+\chi)\nunc 
+ \mathcal{O}\left(a^2k^2/\nunc\right).  \label{diffAs}
\end{equation}
The eigenvectors are 
\begin{equation}
  \begin{split}\label{Xflow}
  f_x=&\left(-1,0,0,-\frac{i c^2 k \sin \theta }{\nunc  (\chi +2)},\chi ,0,0,\frac{2 i c^2 k \chi  \sin \theta }{\nunc  (\chi
   +2)},-\frac{i a k \chi }{\nunc  (\chi +1)},0\right)+\mathcal{O}(\nunc^{-2}),\\[8pt]
   f_y=&\left(0,-1,0,0,0,\chi ,0,0,0,-\frac{i a k \chi  \cos \theta }{\nunc  (\chi +1)}\right)+\mathcal{O}(\nunc^{-2}),\\[8pt]
    f_z=&\left(0,0,-1,-\frac{i c^2 k \cos \theta }{\nunc  (\chi +2)},0,0,\chi ,\frac{2 i c^2 k \chi  \cos \theta }{\nunc  (\chi
   +2)},0,0\right) +\mathcal{O}(\nunc^{-2}),
  \end{split}
 \end{equation}
 corresponding to flows in the $x$, $y$ and $z$ directions respectively. To leading order, all three of these modes have oppositely directed velocities with ratio $|\boldvc|/|\boldvn|=\chi$ and no thermal or magnetic components, so their energy is purely kinetic to this order. All other eigenmodes have near-equal velocities in the highly collisional regime. The rapid decay of these flow eigenmodes represents the decay of the flow differentials.

\paragraph{Isobaric mode slow decay}
\begin{equation}
\omega_\text{pr}=-\frac{2\,i \,c^2k^2(1+\chi)}{\nunc(2+\chi)^2} +\mathcal{O}\left((c\,k/\nunc)^3\right) c\,k.   \label{entAs}
\end{equation}
The eigenvector is
\begin{equation}
\left(\frac{i k \sin \theta }{\nunc },0,-\frac{i k \cos \theta  \left(\chi\tan^2\theta-2
   \right)}{\nunc  (\chi +2)},-1,0,0,-\frac{i k \chi  \sec \theta }{\nunc  (\chi +2)},\chi ,0,0\right)
 +\mathcal{O}(\nunc^{-2}).
 \end{equation}
This mode balances the pressures on the neutrals and charges, $p_\mathrm{c}+p_\mathrm{n}=0$, with small $\mathcal{O}(\nunc^{-1})$ velocity ($u$ and $w$) and magnetic ($b_\perp$) perturbations. Hence it is stationary apart from the slow diffusive decay. \\
\end{subequations}

To the order shown, the three classic MHD waves inherit a slow temporal decay rate of order $1/\nunc$, even for $\chi\gg1$.
The decay rate of the erstwhile stationary isobaric mode is even slower, of order $1/(\chi\nunc)$ in the weakly ionized regime. In stark contrast though, the three modes representing an imbalance between $\boldvc$ and $\boldvn$ decay extremely rapidly, with rate exactly $\nunc+\nucn$ independent of ionization fraction.  For Alfv\'en waves, this is in accord with the analysis of Section 4.1.3 of \citet{SolCarBal13aa}.

The four modes with zero real component of eigenfrequency are called ``entropy modes'' by \citet{SolCarBal13gq}, and ``vortex modes'' by \citet{ZaqKhoRuc11aa} based on the natures of zero-frequency modes in 1F plasmas. However, in the 2F context, we prefer the usage introduced above. Only the isobaric mode has any thermal signature, and so could reasonably be called an ``entropy'' mode, but in view of its pressure balance character, we adopt ``isobaric'' as a more precise descriptor. The flow differential modes are most notable for their discrepant velocities, and so are best named accordingly, rather than with reference to the unrelated vorticity $i\,\k\vcross\boldv$.

\subsection{Cutoffs}
From the exact expression for the Alfv\'en eigenfrequency, it can be shown that there are Alfv\'en cutoff wavenumbers at
\begin{equation}\label{cutoff}
\begin{split}
k_{\text{A,c}}^\pm &=\frac{\nunc\, |\sec\theta|}{8 (\chi +1)a}\,\sqrt{\chi^2+20\chi-8\mp(\chi-8)^{3/2}\chi^{1/2} }\\[4pt]
&= \frac{2\sqrt{2}\,\nunc\,|\sec\theta|}{a}\,\frac{1+\chi}{\sqrt{\chi^2+20\chi-8\pm (\chi-8)^{3/2}\chi^{1/2}}}\\[8pt]
& = 
  \begin{cases}\displaystyle
    \frac{2 \nunc |\sec\theta |}{a}\left(1-\frac{1}{\chi}+\mathcal{O}\left(\chi^{-3}\right)\right) & \text{for ``$+$'' sign,}\\[12pt]
    \displaystyle
    \frac{\nunc |\sec\theta |\sqrt{\chi}}{2a}\left(1+\frac{3}{2\chi} +\mathcal{O}\left(\chi^{-2}\right)\right) & \text{for ``$-$'' sign,}
  \end{cases}
\quad \text{for $\chi\gg1$}.
\end{split}
\end{equation}
The second line here accords with Equation (20) of \citet{SolCarBal13aa} (for the case $\theta=0$), noting that their $c_\text{A}$ is our $a_\text{c}=B/\sqrt{\mu\rhoc}=a\sqrt{1+\chi}$, the charges-only Alfv\'en speed. As they note, there is a real cutoff wavenumber interval only if $\chi>8$. For reference, although $\chi\gg1$ in the low solar atmosphere, it is only about 1 at the top of the chromosphere. The Alfv\'en eigenfrequency is pure imaginary (evanescent) for $k_{\text{A,c}}^+ \le k \le k_{\text{A,c}}^-$. The slow wave is subject to similar cutoffs \citep{SolCarBal13gq}. However, these Alfv\'en and slow cutoffs are at very high wavenumbers and associated frequencies in the solar context, so are of limited practical importance. They are illustrated in Figure \ref{fig:freqs560}.

\subsection{Section Summary}\label{sec:mathsumm}
In summary, the full solution of the wave equations takes the form
\begin{equation}\label{X}
\X(t) =\sum_{m=1}^{10} C_m\X_{\!m}\, e^{-i\,\omega_m t},
\end{equation}
where the $\omega_j$ are the complex eigenfrequencies and the $\X_{\!m}$ are the corresponding eigenvectors. In matrix form, this is
\begin{equation}\label{Xmatrix}
\X(t) =\P e^{-i\,\Omega\, t} \C 
\end{equation}
where $\Omega=\diag[\omega_1,\ldots,\omega_{10}]$, $\C=(C_1,\ldots,C_{10})^T$, and $\P$ is the matrix of eigenvectors of $\M$ arranged as columns consistently with the eigenvalues.

In all cases of interest, the three flow differential modes (arbitrarily $m=5$, 6 and 7) have large negative imaginary eigenfrequencies, and hence disappear almost instantaneously. For all intents and purposes therefore, only the remaining seven eigenvalues and their eigenvectors are significant. We saw analytically in Equations (\ref{omegaAs}) and shall see graphically in Figure \ref{fig:EigEnergy} that these exhibit only minor drift between neutrals and charges.


\section{Energy and Flux} \label{sec:Energy}
It is conventional to introduce quadratic expressions for wave-energy density and wave-energy-density-flux in linear wave studies. Linear energy contributions such as $\B\vdot\b/\mu$ for the magnetic energy average to zero over a period and are ignored. Extending the standard \citet{Eck60aa} process to the 2F equations (\ref{basiceqns}), it is easily shown \citep{CalGom23aa} that
\begin{equation}\label{Eckart}
\pderiv{E}{t} + \Div\F=-\acn\,\left|\boldvn-\boldvc\right|^2,
\end{equation}
where
$E = \rhon |\boldvn|^2/2 + p_{\mathrm{n}}^2/(2\rhon \cn^2) + \rhoc |\boldvc|^2/2 + p_{\mathrm{c}}^2/(2\rhoc \cc^2) +|\b|^2/2\mu$
may be interpreted as a quadratic energy density, and
$\F =  p_{\mathrm{n}}\boldvn + p_{\mathrm{c}} \boldvc -\mu^{-1} \left(\boldvc\vcross\B\right)\vcross\b$
is the corresponding energy density flux. The right hand side represents collisional loss. The energy density consists of respectively the kinetic energy of the neutrals, the thermal energy of the neutrals, the kinetic energy of the charges, the thermal energy of the charges and the magnetic energy. The flux consists of the rate of working by the neutrals and charges pressure perturbations and the Poynting flux.

When a complex representation of the variables is in use, the quadratic terms in $E$ are replaced by their absolute values, so
\begin{equation}
E = \half\rhon |\boldvn|^2 + \frac{|p_{\mathrm{n}}|^2}{2\rhon \cn^2} + \half\rhoc |\boldvc|^2 + \frac{|p_{\mathrm{c}}|^2}{2\rhoc \cc^2} +\frac{|\b|^2}{2\mu},
 \label{E}
\end{equation}
and the flux becomes
\begin{equation}
\F = \re\left( p_{\mathrm{n}}^*\boldvn + p_{\mathrm{c}}^* \boldvc - \frac{1}{\mu} \left(\boldvc\vcross\B\right)\vcross\b^*\right).
\label{F}
\end{equation}
Any particular initialization of oscillations corresponds to a particular partitioning of energy into the ten eigenmodes.

The quadratic wave energy density may be expressed as an Hermitian form in terms of the spectral coefficients $\C$:
\begin{equation}\label{EC}
E=\rhon\, \C^\dagger \Psi(t) \C.
\end{equation}
Similarly, the vertical wave energy flux is
\begin{equation}\label{FC}
F_z=  \half\rhon \C^\dagger \Xi(t)\, \C.
\end{equation}
The Hermitian matrices $\Psi$ and $\Xi$ are described in Appendix \ref{app:EF}. Of course, the flux in any other direction can be calculated similarly, but $F_z$ will serve to illustrate the point.


\section{Initializing Alfv\'en Waves on the Charges Alone}  \label{sec:Alf}
Because of its intrinsic interest as well as its relative simplicity, we now look at the Alfv\'en case, polarized in the $y$-direction, but initiate it in three contrasting ways: (i) with an initial velocity $v_\text{c}$ only; (ii) with an initial magnetic perturbation $b_y$ only; and (iii) with a fully formed Alfv\'en wave on the charges alone. By way of exposition, a further whole-plasma kinetic excitation is also briefly discussed: (iv) initiation via $v_\text{n}=v_\text{c}$ only.

\begin{subequations} \label{AlfInit}
{ \renewcommand{\labelenumi}{(\roman{enumi})}
\begin{enumerate}
\item Ignoring terms of order $\nunc^{-1}$ in the eigenvectors set out above, the velocity-only initiation proposed by \citet{VraPoePan08aa} is constructed uniquely from our eigenvectors as
\begin{equation}\color{black}
\X(0)=(0,0,0,0,0,1,0,0,0,0)=\half(A_- -A_+ +2f_y)/(\chi+1).
\end{equation}
This consists initially of two oppositely directed full-plasma Alfv\'en waves and the $y$-polarized flow differential mode that decays almost instantly. Based on Equations (\ref{EPhi}) and (\ref{Phi}), the total energy density at time $t=0$ associated with unit velocity is $E(0)=\rhon/(2\chi)=\rhoc/2$, as expected. Almost instantly, the $f_y$ component decays due to collisions, leaving only the two oppositely directed Alfv\'en waves, each with energy density $\color{black}E_\pm=\frac{1}{8}\rhoc/ (\chi+1)$, which is negligible in comparison for $\chi\gg1$. The corresponding wave energy fluxes $a\,E_+$ and $-a\,E_-$ directed along the equilibrium magnetic field are similarly small. This is very inefficient.

\item Alternatively, initializing the wave with a magnetic perturbation only, 
\begin{equation}
\X(0)=(0,0,0,0,0,0,0,0,0,1/\sqrt{\chi+1}) =(A_++A_-)/(2\sqrt{\chi+1})
\end{equation}
similarly injects energy $E(0)=\rhoc/2$. However now the modal decomposition contains no decaying flow differential component to leading order. The two energy densities are $E_\pm=\rhoc/4$, equally splitting the energy flux $\pm \,\rhoc a/4$ in the two directions. This is very efficient.

\item Next, we suppose that a fully developed Alfv\'en wave is placed on the charges only at $t=0$, as if the charges and neutrals were collisionally disconnected on $t<0$ with collisions turning on discontinuously at $t=0$. Then 
\begin{equation}
\begin{split}
\X(0) &=\frac{1}{\sqrt{2}}(0,0,0,0,0,-1,0,0,0,(\chi+1)^{-1/2})\\
&=\left\{\half\left[1+(\chi+1)^{-1/2}\right]A_+ + \half\left[1-(\chi+1)^{-1/2}\right]A_- - (\chi+1)^{-1/2} f_y\right\}/(2(\chi+1))^{1/2},
\end{split}
\end{equation}
with total energy $E(0)=\rhoc/2$. Once again, the flow differential mode $f_y$ vanishes rapidly, this time leaving slightly unbalanced Alfv\'en waves propagating in the two field-aligned directions. The respective energy densities are $E_\pm=\rhoc\left(\sqrt{\chi+1}\pm1\right)^2/8(\chi+1)$ for the Alfv\'en waves and $E_y=\rhoc \chi/4(\chi+1)$ for the flow differential mode. In the $\chi\gg1$ regime, these are asymptotically $E_\pm\sim\rhoc/8$, with a small excess at higher order in the prograde Alfv\'en wave compared to the retrograde one, and $E_y\sim \rhoc/4$. So, in effect, this charges-only Alfv\'en excitation rapidly loses half its energy to collisions and initiates oppositely directed Alfv\'en waves each with a quarter of the total energy, and carrying fluxes $\pm \rhoc a/8$ in each direction. Despite the imposed directionality of the original Alfv\'en wave on the charges, the resulting Alfv\'en waves on the full plasma are nearly balanced in the two directions.  This case is 50\% efficient in generating Alfv\'en waves.

\item Finally, we mention the whole-plasma excitation case where the species velocities $v_\text{n}=v_\text{c}$ are initiated together, as envisaged explicitly in Section 6 of \citet{SolCarBal13aa} and implicitly by \citet{TsaSteKop11aa}. Then  
\begin{equation}
\X(0)=(0,-1/\sqrt{\chi+1},0,0,0,-1/\sqrt{\chi+1},0,0,0,0)=\left(A_+ - A_-\right)/(2\sqrt{\chi+1}).
\end{equation}
This again places no energy on the flow differential modes, with almost identical consequences to case (ii): the oppositely directed Alfv\'en waves are generated equally, with no collisional energy loss.

\end{enumerate}}
\end{subequations}

Overall, generation of Alfv\'en waves via the charges alone can range from highly inefficient (case i) to highly efficient (case ii) and in between (case iii). The distinguishing feature is the energy initially placed in the flow differential mode, which is rapidly damped and its energy lost to thermalization. Cases (ii) and (iv) suffer no such losses to leading order in the inverse collision frequency, and so are highly efficient.

Of the three scenarios discussed, case (i), excitation via $v_\text{c}$ only, is the least plausible as well as the least efficient. It is difficult to envisage a mechanism that would inject energy into the charges plasma velocity only and not the magnetic field $b_y$. Specifically, a purely mechanical mechanism could not distinguish between the charges and the neutrals, and an electrical impulse would generate a magnetic perturbation via Faraday induction. 

A similar analytical analysis could be performed for the magneto-acoustic waves, though with greater algebraic complexity. However, we choose to proceed numerically. In the following section Alfv\'en as well as the analogous magneto-acoustic wave excitations are explored via computationally-derived eigenvalues and eigenvectors at three representative levels in the low solar atmosphere for charges-only initiations akin to case (iii) above.

\section{Numerical Results: Three Heights} \label{sec:res}
\begin{figure}[tbthp]
\begin{center}
\includegraphics[width=.85\textwidth]{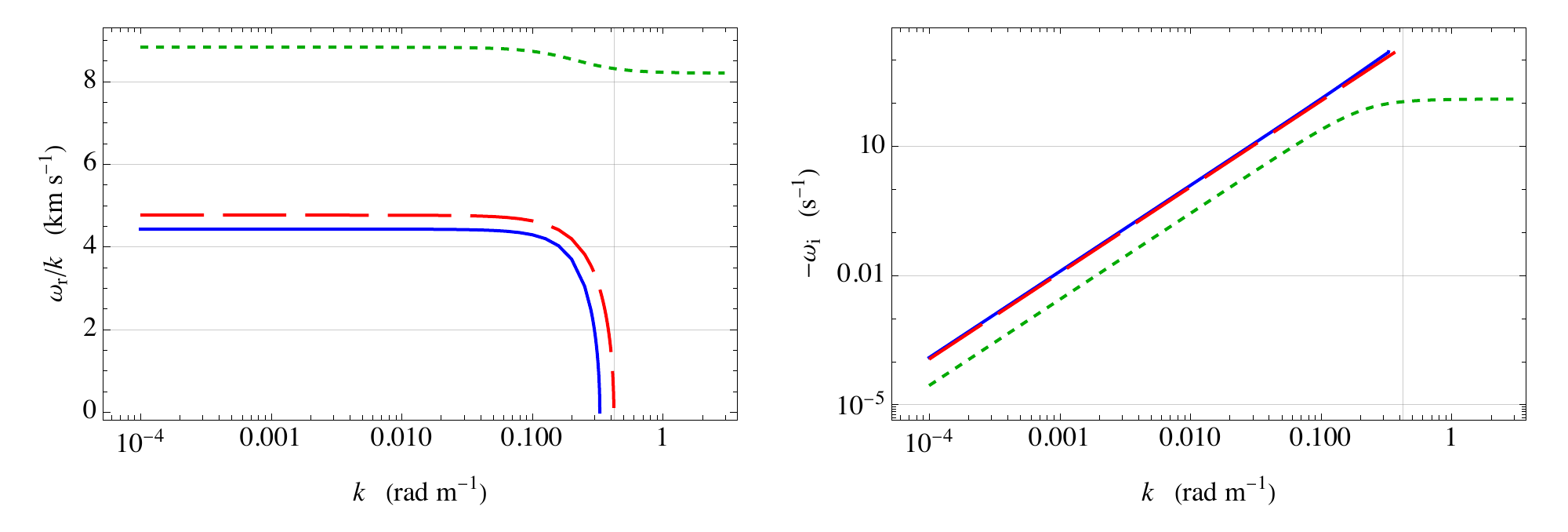}
\caption{Phase speed $\omega_\text{r}/k$ (\kps, log-linear, left) and decay rate $|\omega_\text{i}|$ ($\rm s^{-1}$, log-log, right) plotted against wavenumber $k$ (rad $\rm m^{-1}$) for the temperature minimum model $h=560$ km with $\theta=30^\circ$. For comparison, the specific wavenumber $k_1=0.0063$ referenced in Table \ref{tab:models} and various examples below is comfortably short of the cutoffs and therefore in the asymptotic regime of Equations (\ref{omegaAs}). Blue: slow wave; red long-dashed: Alfv\'en wave; green short-dashed: fast wave. The vertical line indicates the Alfv\'en cutoff $k_{\text{A,c}}^+$.}
\label{fig:freqs560}
\end{center}
\end{figure}

Consider three heights in the solar atmosphere; lower photosphere ($h=0$ km), upper photosphere ($h=250$ km) and temperature minimum ($h=560$ km), with representative sound speed $c$, Alfv\'en speed $a$, ionization ratio $\chi$ and collision frequency $\nunc$ set out in Table \ref{tab:models}. The complex eigenfrequencies $\omega_\text{r}+i\, \omega_\text{i}$ of the slow, Alfv\'en and fast waves are plotted against $k$ in Figure \ref{fig:freqs560} for the temperature minimum $h=560$ km model. The first Alfv\'en cutoff, $k_{\text{A,c}}^+$, and its similar slow counterpart, are apparent. They accurately match the asymptotic formulae given in Equations (\ref{omegaAs}) until the cutoffs start to intrude.

\begin{splitdeluxetable*}{lcccccBcc|cc|cc|cc}
\tablecaption{Atmospheric Models and Wave Eigenfrequencies\label{tab:models}}
\tablehead{
&\colhead{$h$} & \colhead{$c$ } & \colhead{$a$} & \colhead{$\chi$} & \colhead{$\nunc$ } & \colhead{flow $\tau$} & \colhead{isobaric $\tau$} &
\multicolumn{2}{c}{slow} & \multicolumn{2}{c}{Alfv\'en} & \multicolumn{2}{c}{fast} \\
\cline{9-14}
\colhead{} & \colhead{(km)} &\colhead{(\kps)} &\colhead{(\kps)} &\colhead{} &\colhead{($\rm s^{-1}$)} & \colhead{(s)} & \colhead{(s)} & \colhead{Freq (Hz)} & \colhead{$\tau$ (s) }& \colhead{Freq (Hz)} & \colhead{$\tau$ (s)} & \colhead{Freq (Hz)} & \colhead{$\tau$ (s)}
}
\startdata
photospheric base &0 & 10 & 0.63 & 1400 & $7\times 10^5$ & $1.0\times10^{-9}$ & $1.2\times10^5$ & 0.5453 & $9 \times10^4$ &0.5456 & $1.2\times10^5$ & 10 & $6\times10^7$\\
upper photosphere & 250 & 8.6 & 1.4 & 11000 & 1600 & $5.7\times10^{-9}$ & $3.0\times10^4$ & 1.208 & 416 & 1.212 & 551 & 8.6 & $6\times10^4$\\
temperature minimum & 560 & 8.2 & 5.5 & 10500 & 1000 & $9.5\times10^{-8}$ & $2.0\times10^3$& 4.42 & 2.0 & 4.76 & 2.2 & 8.8 & 9.2\\
\enddata
\tablecomments{Representative atmospheric parameters at the photospheric base $h=0$, upper photosphere $h=250$ km and temperature minimum $h=560$ km drawn roughly from Figures 1 and 2 of \cite{CalGom23aa}, adapted for a hydrogen atmosphere from Model C7 of \cite{AvrLoe08aa} with 100 G base magnetic field and a 600 km magnetic scale height fall-off. In the second block, frequencies $\re\omega/2\pi$ (Hz) and decay times $\tau=1/|\im\omega|$ of the slow, Alfv\'en and fast waves in these atmospheres with $k=k_1$ (1 km wavelength) and $\theta=30^\circ$ are listed, as well as the decay times of the $|\boldvn-\boldvc|$ neutrals-charges flow differential and the isobaric mode.}
\end{splitdeluxetable*}

For concreteness, we primarily examine these models with the fiducial wavenumber $k=k_1=2\pi/10^3=0.0063$ $\rm rad\,m^{-1}$, corresponding to a wavelength of 1 km, and $\theta=30^\circ$. A non-zero propagation angle $\theta$ removes any ambiguity between Alfv\'en and slow waves. With these values, the corresponding numerically calculated frequencies $\omega_\text{r}/2\pi$ (Hz) and decay times $\tau=1/\omega_\text{i}$ are also set out in Table \ref{tab:models}. They scale with $k$ as set out in Equations (\ref{omegaAs}).

The slow and Alfv\'en modes are not followed beyond their cutoffs. The decay rates of the acoustically dominated fast wave are well below those of the magnetically dominated slow and Alfv\'en waves.


\begin{figure}[p]
\begin{center}
\includegraphics[width=.77\textwidth]{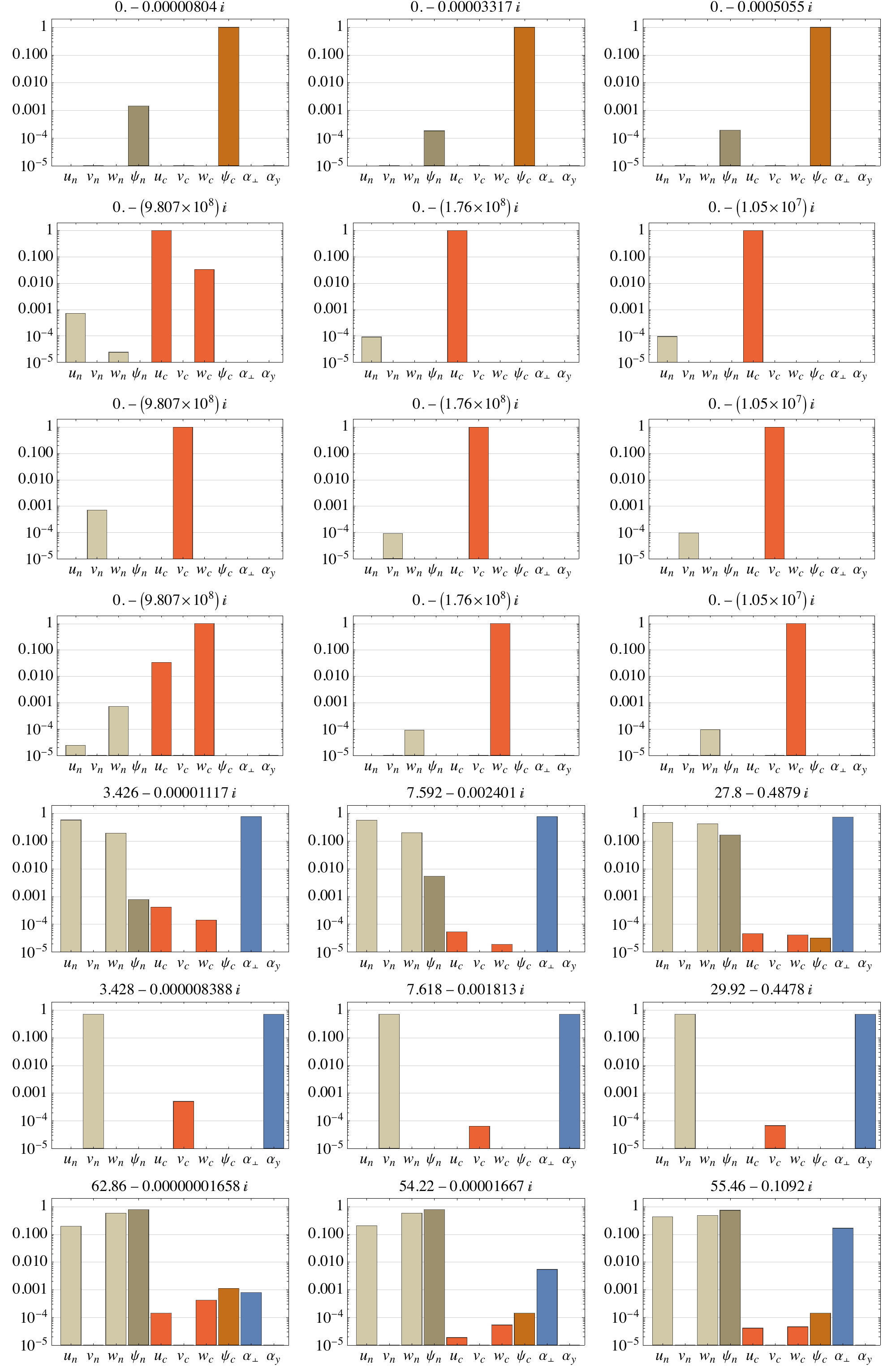}
\caption{Energies in the ten components $(u_{\mathrm{n}},v_{\mathrm{n}},w_{\mathrm{n}}, \psi_{\mathrm{n}},u_{\mathrm{c}},v_{\mathrm{c}},w_{\mathrm{c}}, \psi_{\mathrm{c}}, \alpha_\perp, \alpha_y)$ of the eigenmodes (log scale)} for the three models of Table \ref{tab:models}, $h=0$, 250 and 560 km respectively for the first, second and third columns. In each case, $k=k_1$ and $\theta=30^\circ$. Total energy is normalized to 1, and each mode is labelled with its eigenfrequency. The rows correspond to the different mode types. From top to bottom: isobaric mode; the three flow differential modes; slow; Alfv\'en; fast. The colours correspond to neutrals-kinetic (light fawn); neutrals-thermal (dark fawn); charges-kinetic (red); charges-thermal (red-brown); and magnetic (blue).
\label{fig:EigEnergy}
\end{center}
\end{figure}

In the spirit of \citet{VraPoePan08aa}, the cases where either an Alfv\'en, slow or fast wave is initiated on the \emph{decoupled} charges alone are now explored. This is an extension of Scenario (iii) of Section \ref{sec:Alf}, and represents a sort of thought experiment where we suppose the neutrals and charges are collisionally decoupled on $t<0$, and one of these MHD modes is placed on the charges alone. It is of interest to determine how these charges-only eigenmodes project onto the 10D eigenspace of the coupled system. Collisions are then turned on at $t=0$. 

Figure \ref{fig:EigEnergy}  physically characterizes the eigenmodes by presenting the energy distributions associated with each of the ten components of $\X$ for each of seven eigenmodes of the coupled system (rows) for each $h$ (columns) with $k=k_1$ and $\theta=30^\circ$ (the oppositely directed slow, Alfv\'en and fast modes are omitted as they are the same as for the forward-directed cases). Noting that the ordinate is presented logarithmically, it is seen that:
\begin{itemize}
\item The isobaric mode resides predominantly on the charges. The pressure perturbations in the neutral and charged fluids cancel out, $p_\mathrm{n}+p_\mathrm{c}=0$, allowing a static atmosphere in the strong coupling regime. Therefore, the respective thermal energy densities are in the ratio $2:\chi$, explaining the predominance of the latter at large $\chi$. Kinetic and magnetic energies are negligible in comparison to the therml energy in this mode.
\item In the three flow differential modes, there is a clear difference between $\boldvn$ and $\boldvc$. If they were identical, their respective kinetic energies would be in the ratio $\chi:1$, with more energy on the neutrals, which is certainly not the case. This situation is unsustainable in the presence of inter-species collisions, and hence produces the extremely rapid decay indicated by the eigenfrequencies.
\item The slow modes are dominated by magnetic and neutrals-kinetic energies in the $x$-$z$ plane, with the neutral and charges velocities almost matching.
\item The Alfv\'en modes are polarized in the $y$ direction only, with equipartition between kinetic and magnetic energies. The neutrals and the charges velocities match closely, and hence the neutrals kinetic energy dominates the charges.
\item The fast modes are similar to the slow modes, except that thermal energy plays a much larger role, as is to be expected in predominantly acoustic waves.
\end{itemize}

\begin{figure}[htbp]
\begin{center}
\includegraphics[width=.7\textwidth]{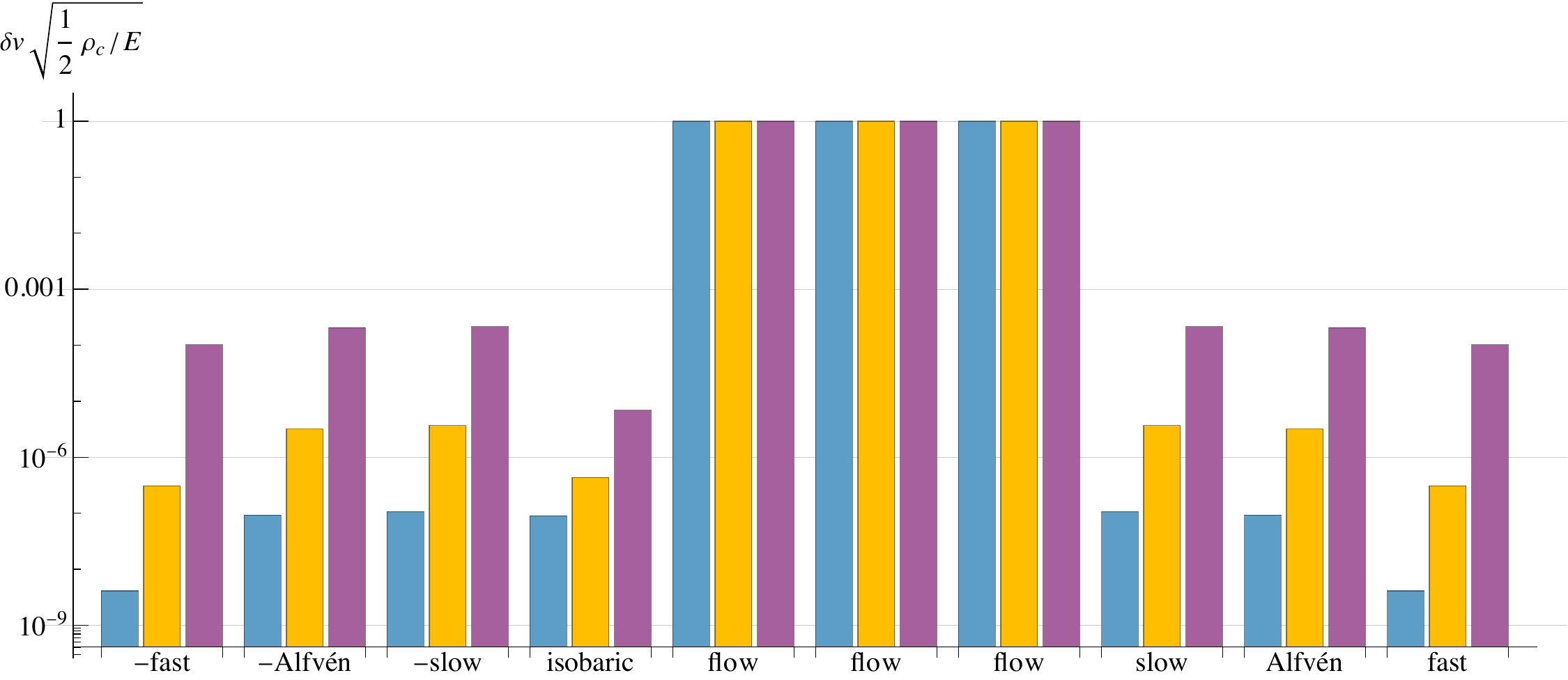}
\caption{Dimensionless velocity differential (logarithmic scale) $|\boldvc-\boldvn|\,\sqrt{\half\rhoc/E}$ associated with each eigenmode (as labelled) for the models at $h=0$ km (blue); $h=250$ km (yellow); and $h=560$ km (black), all with $k=k_1$ and $\theta=30^\circ$. The flow differential $\boldvc-\boldvn$ is dominated by the three extremely rapidly decaying flow differential modes, with the MHD and isobaric modes retaining only small remnant drifts that vanish as $\nunc\to\infty$.}
\label{fig:delta_v}
\end{center}
\end{figure}

Figure \ref{fig:delta_v} shows the distribution of non-dimensionalized $|\boldvc-\boldvn|$ over the ten eigenmodes for each $h$. It is seen that this is dominated by the flow differential modes, with orders of magnitude smaller drift between neutrals and charges on all the other modes. This shows that when these modes are essentially quenched over a few nanoseconds, very little species drift remains, and the wave behaves largely as 1F MHD waves, though with diffusion operating over times of order $\nunc/\omega_\mathrm{r}^2$. From Figure~\ref{fig:EigEnergy}, nearly all energy in the flow differential modes resides in the kinetic energy of the charges flow, explaining why $\sqrt{\half\rhoc|\boldvc-\boldvn|^2/E}\approx1$ for those modes.

\subsection{Projection onto Eigenmodes}

\begin{figure}[htbp]
\begin{center}
\includegraphics[width=\textwidth]{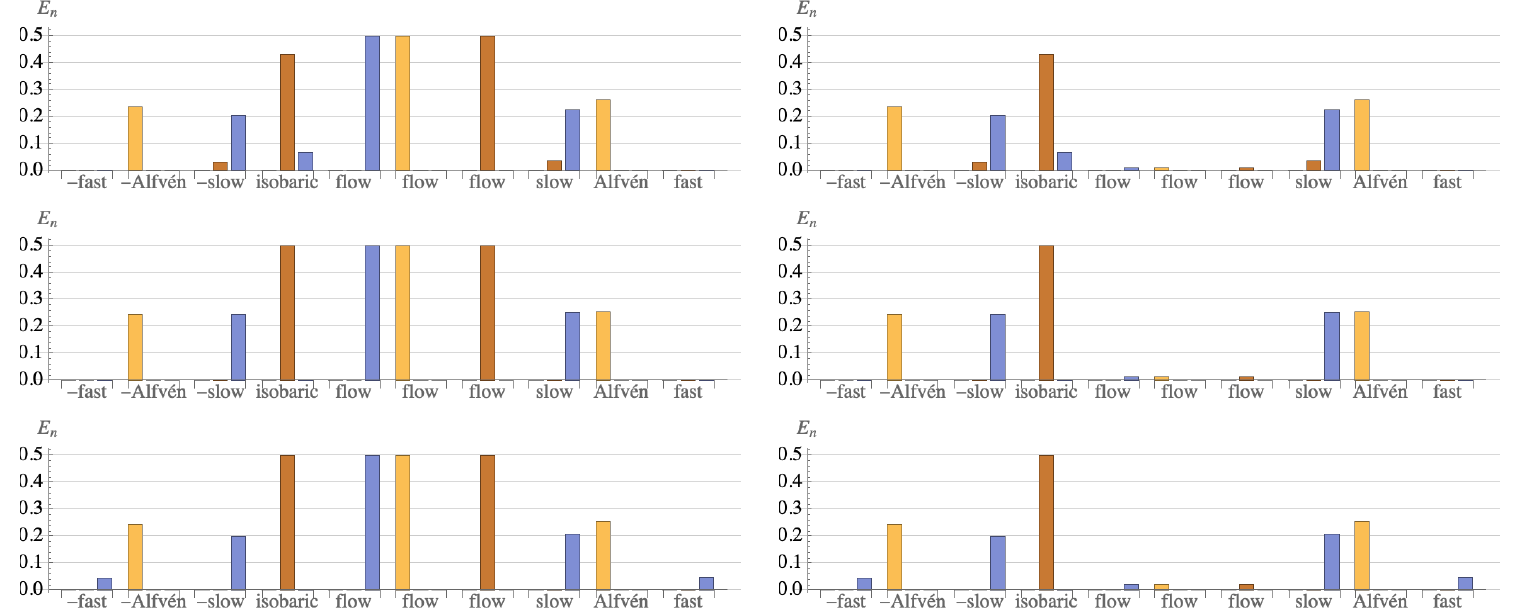}
\caption{Energy in each of the eigenmodes (negatively directed fast, negatively directed Alfv\'en, \ldots, positively directed Alfv\'en, positively directed fast, as labelled) at times $t=0$ (left) and $t=\delta t>0$ (right) for model $h=0$ (top row, $\delta t=2$ ns), $h=250$ km (middle row, $\delta t=10$ ns) and $h=560$ km (bottom, $\delta t=150$ ns) with initiation by Alfv\'en (yellow), slow (brown) and fast (blue) positively directed eigenmodes of the charged fluid only. Total initial energies are normalized to 1. The three flow differential modes have almost disappeared in just nanoseconds, whilst the others have not changed.}
\label{fig:Emodal}
\end{center}
\end{figure}

Figure \ref{fig:Emodal} surveys the modal energy distributions of all three initialization cases at $t=0$ and $t=\delta t$ in the three low-atmosphere models ($\delta t=2$ ns, 10 ns and 150 ns respectively for $h=0$ km, 250 km and 560 km). We see that one flow differential mode accounts for half the energy at $t=0$, but this has essentially disappeared by $t=\delta t$. In essence, the oscillations can be thought to start from this few-nanoseconds state in which the flow differential eigenmodes have been suppressed. 

The modal fluxes in the MHD waves calculated using Equation (\ref{F}) accord perfectly with the energies multiplied by their respective ideal MHD group velocities, 
\begin{equation}
\pderiv{\omega}{k_z}=
  \begin{cases} \displaystyle
\pm \frac{\cos \theta  \left(\Delta  \left(a^2+c^2\right)+2 a^2 c^2 \cos ^2\theta-a^4 -c^4\right)}{\sqrt{2}\, \Delta  \sqrt{a^2+c^2-\Delta }}, & \text{slow}\\[4pt]
   \displaystyle
  \pm  a, & \text{Alfv\'en}\\[4pt]
  \displaystyle
  \pm \frac{\cos \theta  \left(\Delta  \left(a^2+c^2\right)-2 a^2 c^2 \cos ^2\theta +a^4+c^4\right)}{\sqrt{2}\, \Delta  \sqrt{a^2+c^2+\Delta }}, & \text{fast}\\
     \end{cases},
\end{equation}
i.e., $\F=E\,\partial\omega/\partial\k$, and so need not be presented explicitly. However, this serves as a check on the numerics.

An alternative and more realistic scenario is that some driver operates solely on the charges over seconds or minutes. During this period, the near-instantaneous decay of the flow differential modes keeps the neutrals and charges fluid velocities strongly coupled, thereby resulting in energy distributions across the components of $\X$ as set out in the first and the last three rows of Figure \ref{fig:EigEnergy}.  Across the height range, it is seen that:
\begin{itemize}
  \item Initiating with the charges-restricted Alfv\'en wave (yellow) loses half its energy to the flow differential mode, which quickly vanishes, and about a quarter each to the upgoing and downgoing Alfv\'en modes.
  \item Slow wave (acoustic on the charges) initiation (brown) also loses half its energy to the flow differential, with most of the remainder apportioned to the isobaric mode. Very little ends up in the travelling waves.
  \item Fast wave initiation (blue) loses half its energy to the flow differential with around $20-25$\% going to each of the upgoing and downgoing slow waves. A small amount also goes to the two fast waves at $h=560$ km. Recalling that the fast wave on the charges is primarily magnetic, due to the large $\ac$, it was to be expected that it would primarily drive the slow waves (also magnetic) of the coupled plasma.
\end{itemize}
Overall, Alfv\'en and slow waves (the two magnetic waves) are quite efficiently excited and are near-symmetric in direction, whilst generation of the acoustically dominated fast wave is very inefficient for all three drivers. 

Initiating with $b_\perp$ or $b_y$ alone (not shown), and no velocity, just splits the energy into two equal but oppositely directed slow or Alfv\'en modes respectively, \emph{\`a la} d'Alembert. This was described analytically for Alfv\'en waves in Section \ref{sec:Alf}. There is no flow differential at any stage, and hence no energy loss, so this initiation is particularly efficient.  Half of the initialization energy propagates in each direction.

Although there is quite efficient full-plasma wave generation in this charges-only wave initiation scenario, Figure \ref{fig:Emodal} explains why \emph{net} flux is still very small. It is \emph{not} because negligible Alfv\'en (or slow or fast) wave is excited. Instead, despite the initial state being an upward (positive $z$-direction) MHD eigenmode on the charges-only plasma, it projects onto almost equal amplitudes of positively and negatively directed MHD waves in each case. Therefore, when calculating fluxes, these nearly cancel, though each carries up to a quarter of the initial perturbation energy. The initial but fast-disappearing energy in the flow differential modes was never going anywhere anyway, since it had zero $\omega_\mathrm{r}$, and in any case took only half the energy with it.

A real wave excitation region will of course be of finite extent. In most circumstances, it will generate near-equal fluxes of upward and downward propagating MHD waves. In the case of initiation on the charges only, resulting waves on the full plasma carry up to 25\% of the perturbation energy in each direction for the Alfv\'en wave, a little less for slow waves initiated by fast waves, and very little for fast waves. Net flux is therefore very small. However, once they emerge above or below the excitation region, their uni-directional Alfv\'en and slow wave fluxes will be manifest and substantial. It is not correct to say that no significant flux is produced in the \citeauthor{VraPoePan08aa}~scenario.

\pagebreak[4]
\section{Discussion} \label{sec:disc}
For specified wavevector $\k$, the linearized two-fluid collisionally-coupled MHD equations for a weakly ionized plasma imply a ten-dimensional space of solutions, consisting of the familiar six MHD wave modes -- slow, Alfv\'en and fast propagating either forward or backward -- as well as a very slowly decaying stationary isobaric mode and three extremely rapidly decaying (nanoseconds) flow differential mode. The role of these flow modes is to support large velocity differentials between neutrals and charges. They alone carry this responsibility, as the remaining seven eigenmodes intrinsically support only very small differentials.

With hindsight, it could be no other way. It is well known that collisions discharge any significant difference between $\boldvc$ and $\boldvn$ much more rapidly than any other timescale associated with 2F MHD waves, so they must have their own dedicated eigenmodes with eigenfrequencies that have large negative imaginary parts. When $\omega\ll\nunc$, the remaining 2F MHD waves differ only slightly from the standard 1F description, $\nunc\to\infty$, in which $\boldvc=\boldvn$ exactly. The asymptotic eigenfrequency formulae presented in Equations (\ref{omegaAs}) describe these processes in detail. For the Alfv\'en mode specifically, addressed in detail in Section \ref{sec:Alf}, the asymptotic eigenvector corresponding to eigenvalue $\omega_\mathrm{A}$ is given by Equation (\ref{AlfSlip}),
so $\left(v_\mathrm{c}-v_\mathrm{n}\right)/|v_\mathrm{c}| \sim i\,a\, k_z/\nunc$ as $\nunc\to\infty$, showing indeed that the interspecies drift scales as $\nunc^{-1}$ and vanishes in the high-collision limit.

In the cases examined numerically in Section \ref{sec:res}, wave initiation in the form of charges-specific MHD waves results in almost exactly one half of the initial energy residing on the flow differential modes (see Figure \ref{fig:Emodal}), which disappears immediately. The remaining MHD eigenmodes are all roughly symmetrically present in energy, i.e., the forward and backward directed modes of each species have nearly equal energies. This is not an accident. If it were not so, the remaining net flux would not be small, contrary to the very small velocities implied by the vanishing of the flow differential modes. The result is also consistent with our analytical analysis of Section \ref{sec:Alf}.

As in case (ii) of Section \ref{sec:Alf}, an oscillation excited purely via a magnetic perturbation produces no significant flow differential modes and hence essentially no energy loss. The flux still splits into two oppositely directed parts though, now 50:50. This is a very efficient excitation mechanism.

On the other hand, it is also possible in principle to excite the flow differential modes solely, in which case all energy is immediately lost. However, it is difficult to imagine a process that would do this in practice, as it would involve an initial state consisting of one or more of the flow differential eigenmodes given asymptotically in Equation (\ref{Xflow}).

In summary, the ten-dimensional spectral decomposition of the 2F MHD Equations (\ref{basiceqns}) gives considerable insight into how excitation of the charged fluid alone rapidly suppresses interspecies drift via the rapid decay of the flow differential modes. The remaining energy, dependent on the specific initial state, typically finds itself in long-lived MHD waves. This indicates that it is indeed plausible to launch substantial wave flux, especially Alfv\'en wave flux, upward from a weakly ionized photosphere even if the excitation mechanism only directly accesses the charges.

Of course, if excitation is mechanical and equally drives both species in concert, there was never an issue in the first place.

\appendix
\section{Matrix M and its Characteristic Polynomial}\label{app:M}
The fundamental coefficient matrix 
\begin{multline}\label{M}
\M =\\
  \left( \begin{array}{cccc|cccccc}
    -\nunc & 0 & 0 & -i k_x & \nunc & 0 & 0 & 0 & 0 & 0 \\
 0 & -\nunc & 0 & 0 & 0 & \nunc & 0 & 0 & 0 & 0 \\
 0 & 0 & -\nunc & -i k_z & 0 & 0 & \nunc & 0 & 0 & 0 \\
 -\frac{i c^2 (\chi +1) k_x}{\chi +2} & 0 & -\frac{i c^2 (\chi +1) k_z}{\chi +2} & 0 & 0 & 0 & 0 & 0 & 0 & 0 \\[4pt]
 \hline
 \chi  \nunc & 0 & 0 & 0 & -\chi  \nunc & 0 & 0 & -i k_x & i a (\chi +1) k & 0 \\
 0 & \chi  \nunc & 0 & 0 & 0 & -\chi  \nunc & 0 & 0 & 0 & i a (\chi +1) k_z \\
 0 & 0 & \chi  \nunc & 0 & 0 & 0 & -\chi  \nunc & -i k_z & 0 & 0 \\
 0 & 0 & 0 & 0 & -\frac{2 i c^2 (\chi +1) k_x}{\chi +2} & 0 & -\frac{2 i c^2 (\chi +1) k_z}{\chi +2} & 0 & 0 & 0 \\
 0 & 0 & 0 & 0 & i a k & 0 & 0 & 0 & 0 & 0 \\
 0 & 0 & 0 & 0 & 0 & i a k_z & 0 & 0 & 0 & 0 \\
   \end{array}\right),
\end{multline}
where $k_x=k\sin\theta$ and $k_z=k\cos\theta$. Vertical and horizontal dividers have been included to accentuate the neutrals (top left block) and charges (bottom right) submatrices and explicate their collisional couplings (top right and bottom left).

Since the $y$ polarization is decoupled from $x$-$z$, $\M$ can be broken down into separate third and seventh order matrices (it can be expressed in block-diagonal form under the appropriate reordering of rows and columns). Hence,
the tenth order characteristic polynomial (dispersion function) of $\M$ may be factored into a cubic 
\begin{equation}\label{pol3}
\omega ^2 \left(\omega+i\,(\chi+1)  \nunc\right)-a^2 k^2 (\chi +1) \cos ^2\theta  \left(\omega +i\,\nunc\right) =0,
\end{equation}
which captures the two Alfv\'en modes and one flow differential mode (those oscillations polarized in the $y$-direction), and an equation of the seventh order \citep[see also Eq.~(57) of][]{ZaqKhoRuc11aa} that contains the four magneto-acoustic modes, two flow differential modes and the isobaric mode:
\begin{multline}\label{pol7}
\omega\left(\omega^4-(a^2+c^2)k^2\omega^2+a^2c^2k^4\cos^2\theta \right) =\\
\omega\,\tau ^2 \left(
\frac{\omega ^6}{(\chi +1)^2}
-\frac{k^2 \omega ^4 \left(a^2 (\chi +2)+3 c^2\right)}{(\chi +1) (\chi +2)}
+\frac{c^2 k^4 \omega ^2 \left(2 \left(a^2 (\chi +2)+c^2\right)+a^2 (\chi +2)
   \cos 2 \theta \right)}{(\chi +2)^2}
-\frac{2 a^2 c^4 k^6 (\chi +1)   \cos ^2\theta }{(\chi +2)^2}   \right)\\
   +i \,\tau 
   \left(\frac{2  \,\omega ^6}{\chi  +1}
   -\frac{ k^2 \omega ^4 \left(a^2 (\chi +2)^2+c^2 (4 \chi +5)\right)}{(\chi +1) (\chi +2)}
   +\frac{ c^2 k^4 \omega ^2 \left(2 a^2 (\chi +2) \cos 2 \theta +a^2 (\chi +2) (\chi
   +3)+2 c^2 (\chi +1)\right)}{(\chi +2)^2} \right.\\
   \left.
   -\frac{2  a^2 c^4 k^6 (\chi +1) \cos ^2\theta }{(\chi +2)^2}\right),
\end{multline}
where $\tau=1/\nunc$ is the neutrals-charges collision timescale. In the fully coupled limit $\tau\to0$, this dispersion relation reduces to that of the classic 1F magneto-acoustic modes and the stationary isobaric mode $\omega=0$. The full tenth order formulation is retained here for unity of exposition.


\section{Eigensystem Perturbation Analysis}\label{app:pert}
Although the eigenvalues and eigenvectors of $\M$ may be found numerically, the analytic large-$\nunc$  asymptotic formulae set out in Equations (\ref{omegaAs}) are valuable aids to understanding. The process of developing these formulae is not straightforward though. A matrix method is sketched here,
beginning with splitting $\M=\M_0+\nunc\M_1$. Thus
\begin{equation}\label{pertEq}
\left(\M_0+\nunc\M_1\right)\X = \lambda\X.
\end{equation}
Recall that the eigenfrequencies are $\omega=i\,\lambda$.

At large $\nunc$, $\M_1$ plays the leading role. It is only rank-3, but is diagonalizable: $\M_1=\S \Lambda \S^{-1}$, where the columns of $\S$ are the eigenvectors and the diagonal matrix $\Lambda=\diag(-(\chi+1),-(\chi+1),-(\chi+1),0,0,0,0,0,0,0)$ is made up of the eigenvalues. The non-zero entries correspond to the three flow differential modes. (Alternatively, LU decomposition works just as well in that it also collects these modes in only the first three rows of the upper triangular $\U$ matrix.) Then Equation (\ref{pertEq}) may be recast as
\begin{equation}\label{pertEqY}
\left(\Q+\nunc\Lambda\right)\Y = \lambda\Y,
\end{equation}
where $\Q=\S^{-1}\M_0\S$ and $\Y=\S^{-1}\X$.

The required procedure then differs between the flow differential and the other seven modes.

\subsection{Flow Differential Modes}
Substituting $\lambda=\nunc(\lambda_0+\nunc^{-1}\lambda_1+\nunc^{-2}\lambda_2+\cdots)$ and $\Y=\Y_{\!\!0}+\nunc^{-1}\Y_{\!\!1}+\nunc^{-2}\Y_{\!\!2}+\cdots$ into Equation (\ref{pertEqY}) and equating powers of $\nunc$, it is found that
\begin{equation}
  \Lambda\Y_{\!\!0} = \lambda_0\Y_{\!\!0} ,
\end{equation}
 an eigenvalue equation that determines $\lambda_0=-(\chi+1)$ for the three flow differential modes and the corresponding leading order eigenvectors $\Y_{\!\!0}$. 
 
 At the next order, 
\begin{equation}
  \left(\Lambda-\lambda_0\I\right)\Y_{\!\!1}-\lambda_1\Y_{\!\!0}=-\Q\Y_{\!\!0},
\end{equation}
etc. This may be solved for $\lambda_1$ and $\Y_{\!\!1}$ by writing it as an $11\times11$ matrix equation for $(\Y_{\!\!1},\lambda_1)$ when supplemented with a normalization condition. In practice, one component of $\Y_{\!\!1}$ corresponding to a nonzero component of $\Y_{\!\!0}$ is set to zero keeping that component of $\Y$ unchanged under the perturbation. Equations (\ref{Xflow}) result on recovering $\X=\S\Y$. The procedure may be repeated to higher order if required.

No use is made of the remaining seven eigen-solutions, which are better attacked as follows.

\subsection{MHD and Isobaric Modes}
Defining $\J=\diag(0,0,0,1,1,1,1,1,1,1)$ it is clear that $\J\Lambda=0$, and Equation (\ref{pertEqY}) may be used to derive
\begin{equation}\label{pertEqYJ}
\left(\R+\nunc^{-1}\Q\right)\Y = \lambda\left(\J+\nunc^{-1}\I\right)\Y,
\end{equation}
where $\R=\Lambda+\J\Q$.

With $\lambda=\lambda_0+\nunc^{-1}\lambda_1+\nunc^{-2}\lambda_2+\cdots$, and $\Y$ expanded as before,  it follows that
\begin{equation}
  \R\Y_{\!\!0} = \lambda_0\J\Y_{\!\!0} .
\end{equation}
This is a generalized eigenvalue equation from which the leading behaviours of the remaining seven $\lambda$ and $\Y$ may be determined.

At the next order, 
\begin{equation}
  \left(\R-\lambda_1\J\right)\Y_{\!\!1}-\lambda_1\J\Y_{\!\!0}=\left(\lambda_0\I-\Q\right)\Y_{\!\!0},
\end{equation}
which may again be used to solve for $\lambda_1$ and $\Y_{\!\!1}$ subject to a normalization. The remainder of Equations (\ref{omegaAs}) result.

\section{Energy and Flux in Matrix Form}\label{app:EF}
In complex matrix form, we may write
\begin{equation}\label{EPhi}
E=\rhon\,\X^\dagger\Phi\X
\end{equation}
where the dagger indicates conjugate transpose and
\begin{equation} \label{Phi}
\Phi= \half\diag\left[1,1,1,\frac{\chi+2}{c^2(\chi+1)},\frac{1}{\chi},\frac{1}{\chi},\frac{1}{\chi},\frac{\chi+2}{2c^2\chi(\chi+1)},\frac{\chi+1}{\chi},\frac{\chi+1}{\chi}\right].
\end{equation}

In terms of the eigenvector decomposition and coefficient vector $\C$ the energy density is
\begin{equation}
E=\rhon\, \C^\dagger \Psi(t) \C,
\end{equation}
where $\Psi(t)=e^{i\,\Omega^*t}\P^\dagger \Phi \P e^{-i\,\Omega\,t}$, and $\Omega=\diag[\omega_1,\ldots,\omega_{10}]$. By construction, $\Psi$ is Hermitian. In principle, since $\Psi$ is not diagonal, there is cross-talk between the modes. However, in practice, the off-diagonal contributions to $E$ appear in purely imaginary complex conjugate pairs which do not contribute at all to the overall energy as they cancel, and in any case they are entirely negligible in magnitude. For that reason, we may attribute all energy to the ten individual eigenmodes and not interactions between them.

The $z$-component of flux $F_z$ for example may also be written as a quadratic form,
\begin{equation}
F_z=  \half\rhon \X^\dagger \Upsilon \X,
\end{equation}
where the real symmetric (and therefore Hermitian) matrix
\begin{equation}
\Upsilon=
\begin{pmatrix}
 0 & 0 & 0 & 0 & 0 & 0 & 0 & 0 & 0 & 0 \\
 0 & 0 & 0 & 0 & 0 & 0 & 0 & 0 & 0 & 0 \\
 0 & 0 & 0 & 1 & 0 & 0 & 0 & 0 & 0 & 0 \\
 0 & 0 & 1 & 0 & 0 & 0 & 0 & 0 & 0 & 0 \\
 0 & 0 & 0 & 0 & 0 & 0 & 0 & 0 & -\frac{a (\chi +1) \cos \theta }{\chi } & 0 \\
 0 & 0 & 0 & 0 & 0 & 0 & 0 & 0 & 0 & -\frac{a (\chi +1)}{\chi } \\
 0 & 0 & 0 & 0 & 0 & 0 & 0 & \frac{1}{\chi } & 0 & 0 \\
 0 & 0 & 0 & 0 & 0 & 0 & \frac{1}{\chi } & 0 & 0 & 0 \\
 0 & 0 & 0 & 0 & -\frac{a (\chi +1) \cos \theta }{\chi } & 0 & 0 & 0 & 0 & 0 \\
 0 & 0 & 0 & 0 & 0 & -\frac{a (\chi +1)}{\chi } & 0 & 0 & 0 & 0 \\
\end{pmatrix},
\end{equation}
or in terms of $\C$,
\begin{equation}
F_z=  \half\rhon \C^\dagger \Xi\, \C,
\end{equation}
where $\Xi=e^{i\,\Omega^*  t}\,\P^\dagger \Upsilon \P\, e^{-i\,\Omega\,t}$.

Unlike the energy density, the cross-talk in flux between eigenmodes can be substantial in magnitude. However, cross-talk between MHD modes averages to zero over time, as does the interaction between a wave mode and the isobaric mode. The interaction with a flow differential mode of course quickly vanishes. So only the diagonal entries, attributed to slow, Alfv\'en and fast waves, contribute meaningfully.

\subsection*{}\noindent
 The author thanks Elena Khomenko and Martin G\'omez M\'iguez for their very useful comments and suggestions on an initial draft of this paper.

\bibliography{fred}
\bibliographystyle{aasjournal}

\end{document}